\documentclass[onecolumn,aps,a4paper,11pt]{revtex4}

\usepackage{graphicx}% Include figure files
\usepackage{epsfig}
\usepackage{dcolumn}% Align table columns on decimal point
\usepackage{bm}% bold math

\def\Journal#1#2#3#4{{#1} {\bf #2}, #3 (#4)}

\def\NIMA{{\em Nucl. Instrum. Methods} A}

\def\PRL{\em Phys. Rev. Lett.}

\def\be{\begin{equation}}
\def\ee{\end{equation}}
\def\bea{\begin{eqnarray}}
\def\eea{\end{eqnarray}}

\newcommand{\met}{{\not\! E_{\rm T}}  }
\begin{document}
%\vspace*{4cm}
\title{
%\rightline{CDF/PUB/CDF/PUBLIC/10250}
Experimental results on diffraction at CDF}

\author{Michele Gallinaro\footnote{On behalf of the CDF Collaboration}\footnote{Presented at the workshop on ``Forward physics at the LHC'', Biodola, May 27-29, 2010}}

\affiliation{The Rockefeller University (USA) and LIP Lisbon, Portugal}

\begin{abstract}
Diffractive events are studied by means of identification of one or more rapidity gaps and/or a leading antiproton.
Measurements of soft and hard diffractive processes have been performed at the Tevatron $p\bar p$ collider and presented.
%Experimental results on diffraction from the Fermilab Tevatron collider obtained by the CDF experiment are presented.
We report on the diffractive structure function obtained from dijet production 
in the range $0<Q^2<10,000$~GeV$^2$, and on the
$|t|$ distribution in the region $0<|t|<1$~GeV$^2$ for both soft and hard diffractive events up to $Q^2\approx 4,500$~GeV$^2$.
Results on single diffractive W/Z production, forward jets, 
and central exclusive production of both dijets and Z-bosons are also presented.
\end{abstract}
\maketitle

\section{Introduction}

%Diffractive interactions have been traditionally discussed in terms of the ``Pomeron'' and correspond to approximately 25\% of the inelastic collisions at the Tevatron.
Diffraction can be described as an exchange of a combination of quarks and gluons carrying the quantum numbers of the vacuum~\cite{dino}.
As no radiation is expected from such an exchange, diffractive processes are characterized by the presence of large rapidity regions 
not filled with particles (``rapidity gaps'').

%
%Hadronic diffraction has traditionally been treated in the framework of Regge theory.
%In the Regge theory one assumes that the elastic amplitude is mediated by exchange of the so-called Pomeron.
%Diffractive processes are described by the exchange of the ``Pomeron trajectory'', 

%The Regge trajectory corresponding to this singularity is approximately linear, i.e. $\alpha(t)=\alpha_0+\alpha^\prime\cdot t$,
%with parameters $\alpha_0=1.1$ and $\alpha^\prime=0.25$~GeV$^{-2}$, and it is 
%presumed to be formed by a group of particles carrying the quantum numbers of the vacuum.
%This behavior follows from data for elastic and total cross sections.
%However, the linear $t-$dependence of the Pomeron trajectory cannot continue forever at large negative $t$. 
%Indeed, the higher order corrections vanish as powers of the QCD coupling $\alpha_s$, and the Pomeron trajectory $\alpha(t)$ should approach the value corresponding 
%to the Born graph, $\alpha(t)\rightarrow 1$. 
%Indeed, the trajectory seems to level off at large $|t|$ according to data [1].
%UA8 Collaboration, A. Brandt et al., Nucl. Phys. B 514, 3 (1998).
%

%Traditionally discussed in terms of the ``Pomeron'',
%diffraction can be described as an exchange of a combination of quarks and gluons carrying the quantum numbers of the vacuum~\cite{dino}.

At the Fermilab Tevatron collider, proton-antiproton collisions have been used to study diffractive interactions 
in Run~I (1992-1996) at an energy of $\sqrt{s}=1.8$~TeV and continued in Run~II (2003-present) with new and upgraded detectors at $\sqrt{s}=1.96$~TeV.
The goal of the CDF experimental program at the Tevatron is to provide results help decipher
the QCD nature of hadronic diffractive interactions, and to measure exclusive production rates which could be used to 
establish the benchmark for exclusive Higgs production at the Large Hadron Collider (LHC).
The study of diffraction has been performed by tagging events either with a rapidity gap or with a leading hadron.
The experimental apparatus includes a set of forward detectors\cite{fd} that 
extend the rapidity~\cite{rapidity} coverage to the forward region. 
The Miniplug (MP) calorimeters cover the region $3.5<|\eta|<5.1$; the Beam Shower Counters (BSC) surround the beam-pipe 
at various locations and detect particles in the region $5.4<|\eta|<7.4$;
the Roman Pot spectrometer (RPS) tags the leading hadron scattered from the interaction point after 
losing a fractional momentum approximately in the range $0.03<\xi<0.10$.

\section{Diffractive dijet production}

The gluon and quark content of the interacting partons can be investigated by comparing 
single diffractive (SD) and non diffractive (ND) events.
SD events are triggered on a leading anti-proton in the RPS
and at least one jet, while the ND trigger requires at least one jet in the calorimeters.
The ratio of SD to ND dijet production rates ($N_{jj}$) is proportional to the ratio 
of the corresponding structure functions ($F_{jj}$),
$R_{\frac{SD}{ND}}(x, \xi, t)= \frac{N_{jj}^{SD}(x, Q^2, \xi, t)}{N_{jj}(x, Q^2)}
\approx \frac{F_{jj}^{SD}(x, Q^2, \xi, t)}{F_{jj}(x, Q^2)}$,
and can be measured as a function of the Bjorken scaling variable $x\equiv x_{Bj}$\cite{xbj}.
In the ratio, jet energy corrections approximately cancel out, thus avoiding dependence on Monte Carlo (MC) simulation.
Diffractive dijet rates are suppressed by a factor of {\cal O(10)} with respect to expectations based on the proton PDF obtained 
from diffractive deep inelastic scattering at the HERA $ep$ collider~\cite{dino}.
The SD/ND ratios (i.e. gap fractions) of dijets, W, b-quark, $J/\psi$ production are all approximately 1\%, indicating that 
the suppression factor is the same for all processes and it is related to the gap formation.

\begin{figure}[htbp]
\begin{center}
\includegraphics[width=0.49\textwidth]{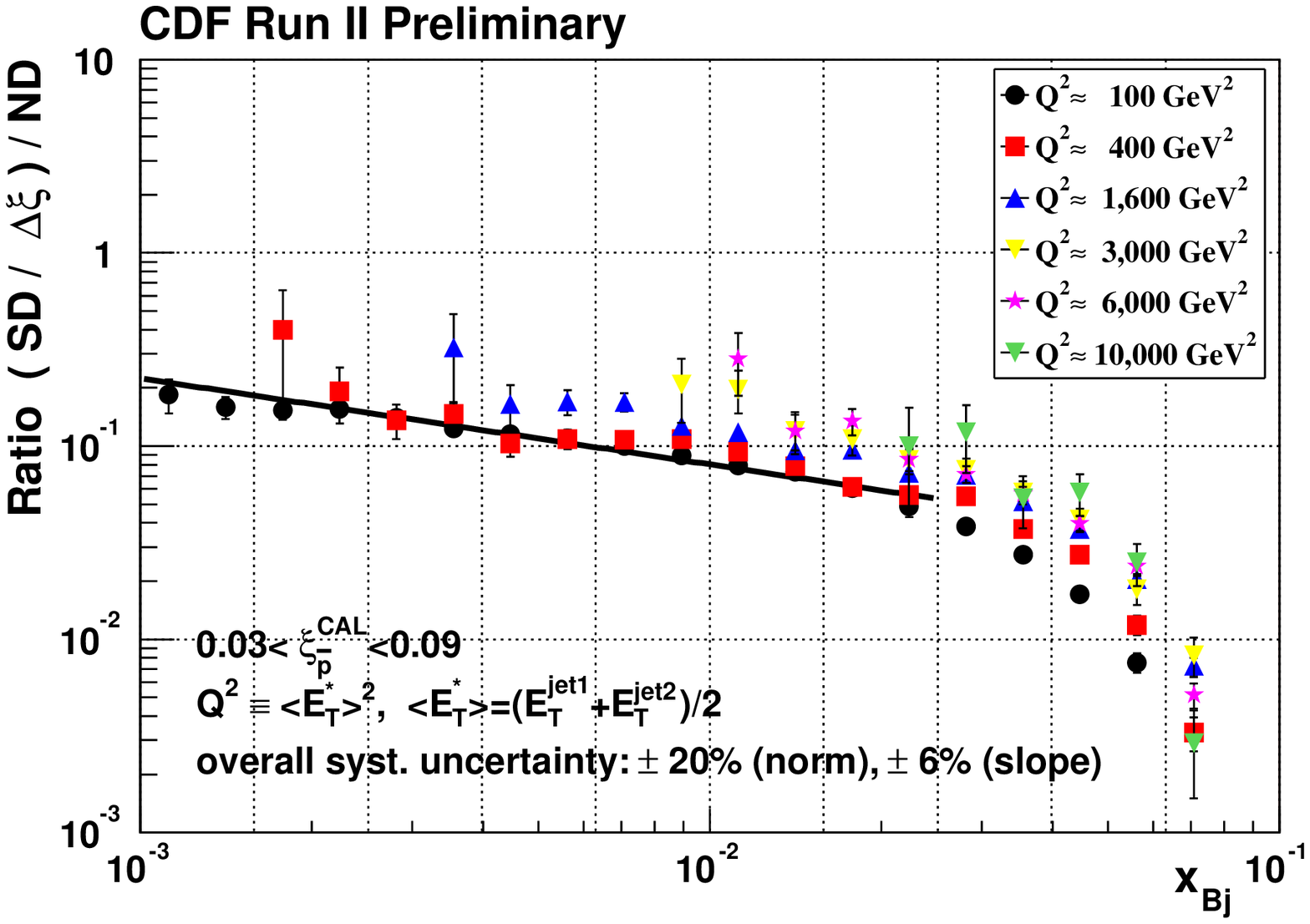}
\includegraphics[width=0.49\textwidth]{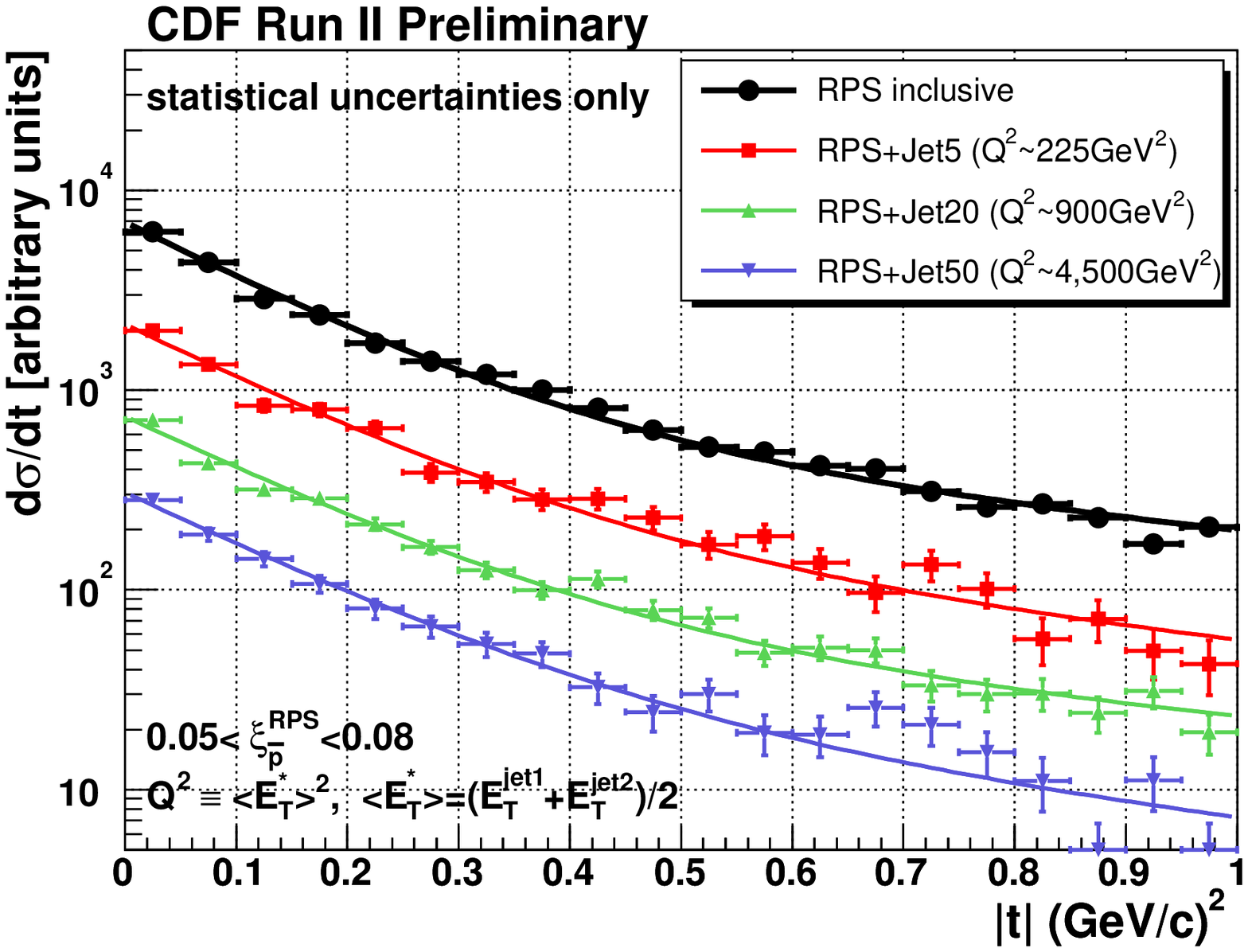}
\caption{
{\em Left}: Ratio of diffractive to non-diffractive dijet event rates
as a function of $x_{Bj}$ (momentum fraction of struck parton in the anti-proton) for different values of $E{_T}^2 \equiv Q^2 $;
{\em Right}: Measured $|t|$-distributions for soft and hard diffractive events.
}
\label{fig:dsf}
\end{center}
\end{figure}

In Run~II, the jet $E_T$ spectrum extends to $E_T^{\rm jet}\approx 100$~GeV, and 
results are consistent with those of Run~I\cite{run1_dsf}, hence confirming a breakdown of factorization. 
Preliminary results indicate that the ratio does not strongly depend on $E_T^2\equiv Q^2$ 
in the range $100<Q^2<10,000$~GeV$^2$ (Fig.~\ref{fig:dsf}, left).
The relative normalization uncertainty cancels out in the ratio, and the results indicate that the $Q^2$ evolution, 
mostly sensitive to the gluon density, is similar for the proton and the Pomeron.
A novel technique~\cite{dis06} to align the RPS is used to measure 
the diffractive dijet cross section
as a function of the $t$-slope in the range up to $Q^2\simeq 4,500$~GeV/c$^2$ (Fig.~\ref{fig:dsf}, right).
The shape of the $t$ distribution does not depend on the $Q^2$ value, in the region $0\le |t| \le 1$~GeV$^2$.
Moreover, the $|t|$ distributions do not show diffractive minima, which could be caused by the interference 
of imaginary and real parts of the interacting partons.

\section{Diffractive W/Z production}

Studies of diffractive production of the W/Z bosons are an additional handle to the understanding of diffractive interactions.
At leading order (LO) diffractive W/Z bosons are produced by a quark interaction in the Pomeron. 
Production through a gluon can take place at NLO, which is suppressed by a factor $\alpha_s$ 
and can be distinguished by the presence of one additional jet.

In Run~I, the CDF experiment measured a diffractive $W$ boson event rate $R_W=1.15\pm0.51$~(stat)$\pm 0.20$~(syst)\%. 
Combining the $R_W$ measurement with the dijet production event rate (which takes place both through quarks and gluons) and with the b-production rate
allows the determination of the gluon fraction carried by the Pomeron which can be estimated to be $54^{+16}_{-14}$\%~\cite{gluonfraction}.

\begin{figure}[htbp]
\begin{center}
\includegraphics[width=0.44\textwidth]{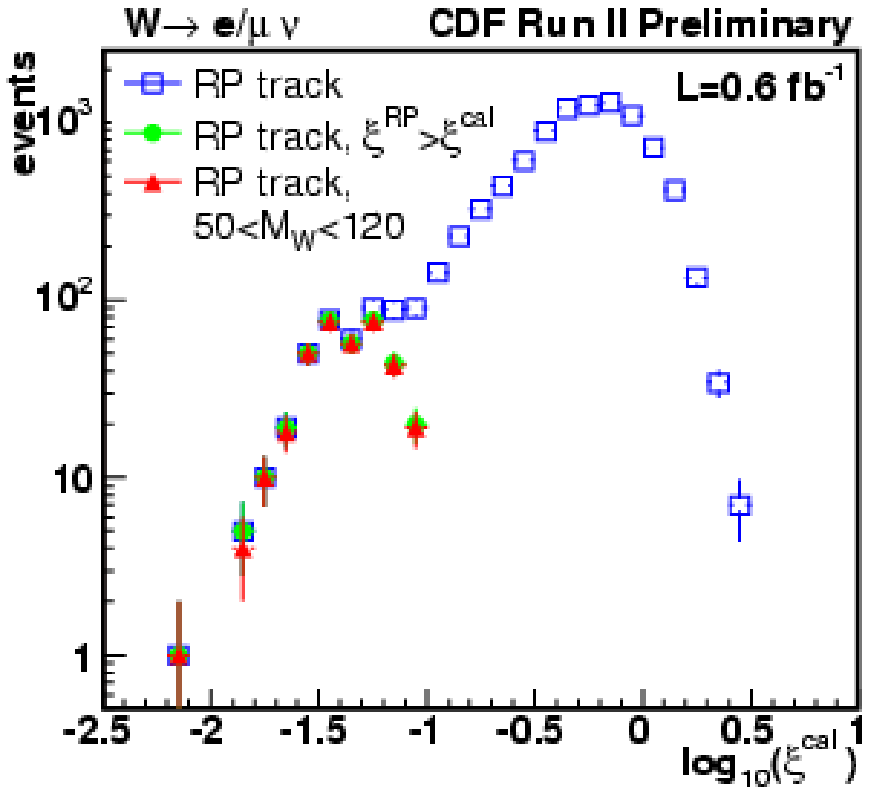}
\includegraphics[width=0.49\textwidth]{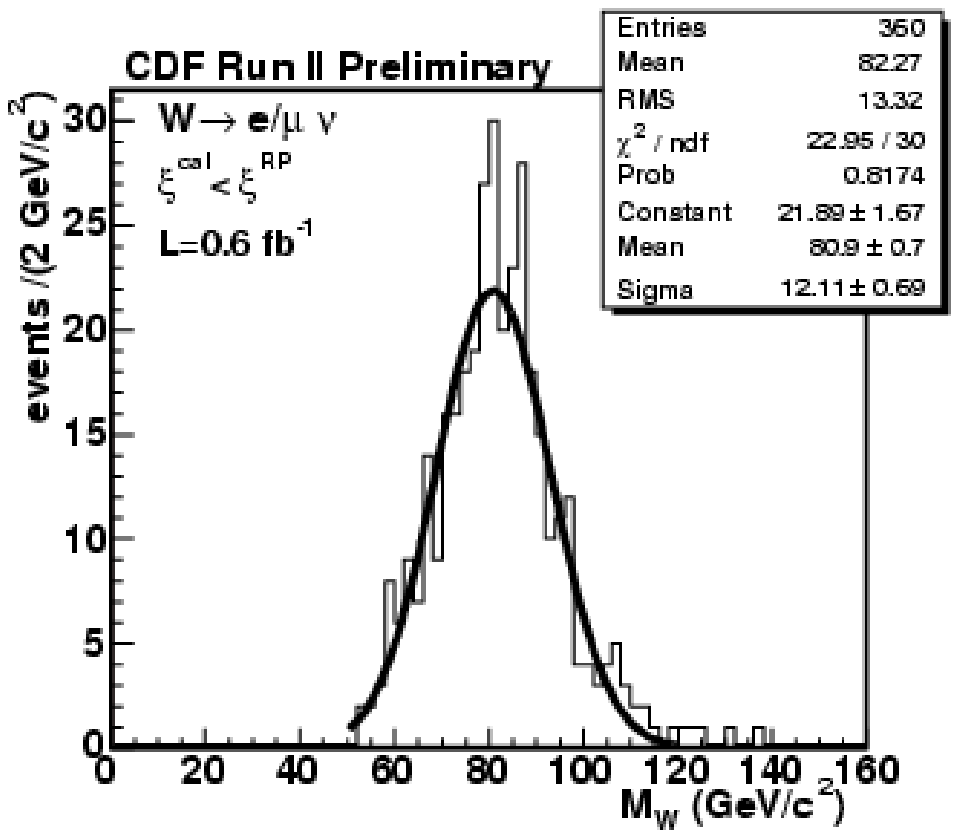}
\caption{Calorimeter $\xi^{cal}$ distribution in $W$ events with a reconstructed Roman Pot track ({\it left}).
Due to the neutrino, $\xi^{cal}<\xi^{RPS}$ is expected. The difference $\xi^{RPS}-\xi^{cal}$ is used to determine the $W$ boson mass ({\it right}).}
\label{fig:diffW}
\end{center}
\end{figure}

In Run~II, the RPS provides an accurate measurement of the 
fractional energy loss ($\xi$) of the leading hadron (Fig.~\ref{fig:diffW}, left), removing the ambiguity of the gap survival probability.
The innovative approach of the analysis takes advantage of the full
$W\rightarrow l\nu$ event kinematics including the neutrino.
The missing transverse energy ($\met$) is calculated as usual from all calorimeter towers, and the neutrino direction (i.e. $\eta_\nu$) 
is obtained from the comparison between the fractional energy loss measured in the Roman Pot spectrometer ($\xi^{RPS}$) 
and the same value estimated from the calorimeters ($\xi^{cal}$): 
$\xi^{RPS}-\xi^{cal}={\met\over\sqrt{s}}\cdot e^{-\eta_\nu}$. 
The reconstructed $W$ mass (Fig.~\ref{fig:diffW}, right) yields $M_W=80.9\pm 0.7$~GeV/c$^2$, in good agreement with the world average value of
$M_W=80.398\pm 0.025$~GeV/c$^2$\cite{pdg}.
After applying the corrections due to the RPS acceptance, trigger and track reconstruction efficiencies, and taking into account the effect of multiple interactions,
both $W$ and $Z$ diffractive event rates are calculated:
$R_W=0.97\pm 0.05 {\rm (stat)}\pm 0.10{\rm (syst)}\%$, and $R_Z$=0.85$\pm$0.20~(stat)$\pm$0.08~(syst)\%~\cite{mary}.

Search for exclusive Z candidate events, i.e. $p\bar p\rightarrow p + Z + \bar p$, has been explored with a null result.
In the SM, the process takes places through photo-production. %, where a virtual photon is radiated from the $p$($\bar p$).
The search requires that nothing else is found in the detector, except the two leptons from $Z\rightarrow ll$.
The method consists in comparing the total energy in the calorimeter ($M_X$) to the dilepton invariant mass ($M_{ll}$). 
Exclusive events are expected to be found on the diagonal $M_X$=$M_{ll}$. Some effects that may artificially change the ``exclusive'' behavior.
By increasing the calorimeter thresholds the value of $M_X$ moves closer to the diagonal $M_X$=$M_{ll}$. 
Because of charge conservation, $W$ bosons cannot be produced exclusively and are used as the control sample. %The diffractive $W$ sample is used as the control sample.
Additional control of the background is performed by looking at ``empty crossings'', where no tracks are reconstructed and calorimeter noise is the dominant effect.
No exclusive candidate are found in the data.

\section{Forward jets}

\begin{figure}[htbp]
\begin{center}
\includegraphics[width=0.49\textwidth]{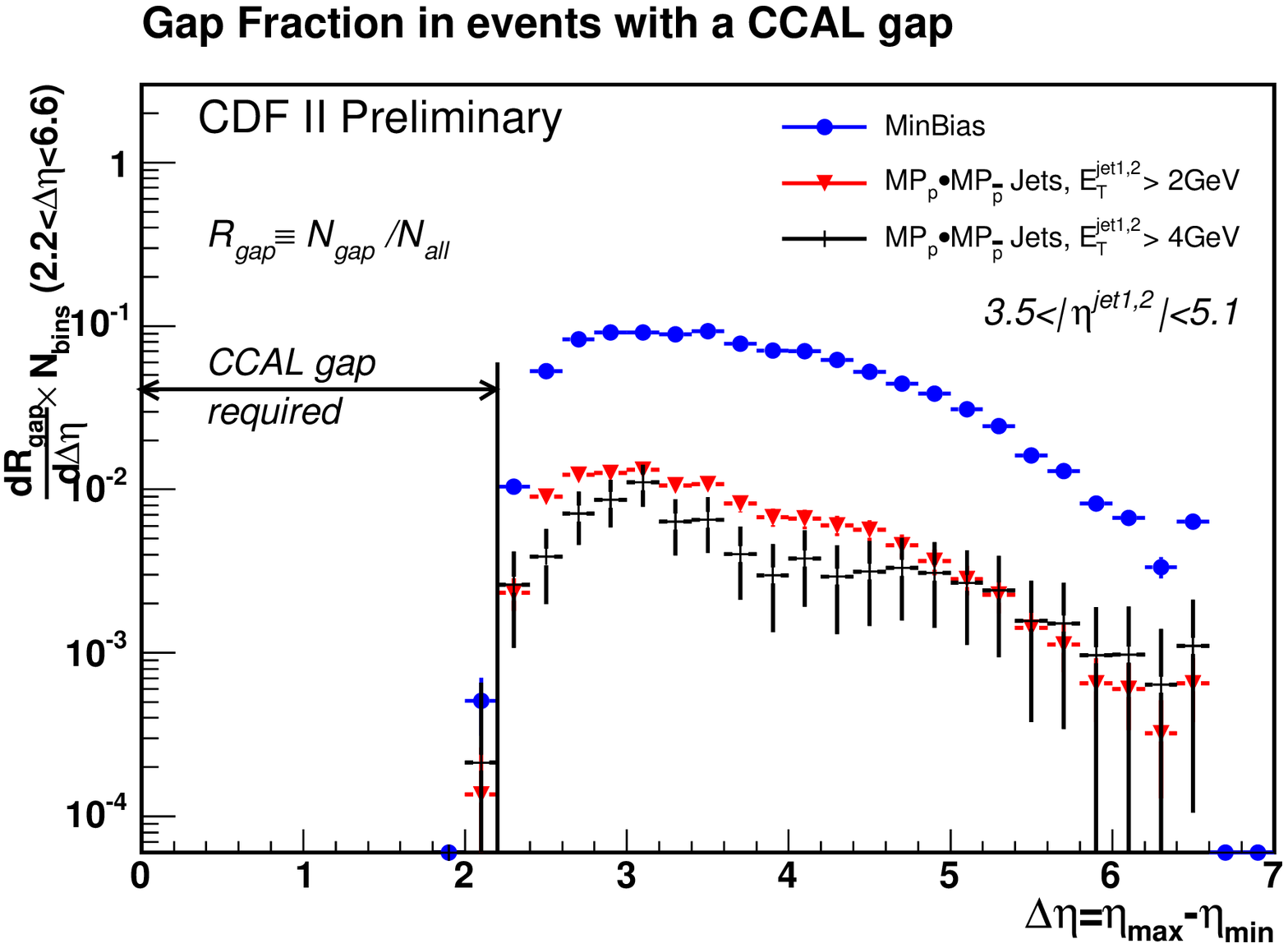}
\includegraphics[width=0.49\textwidth]{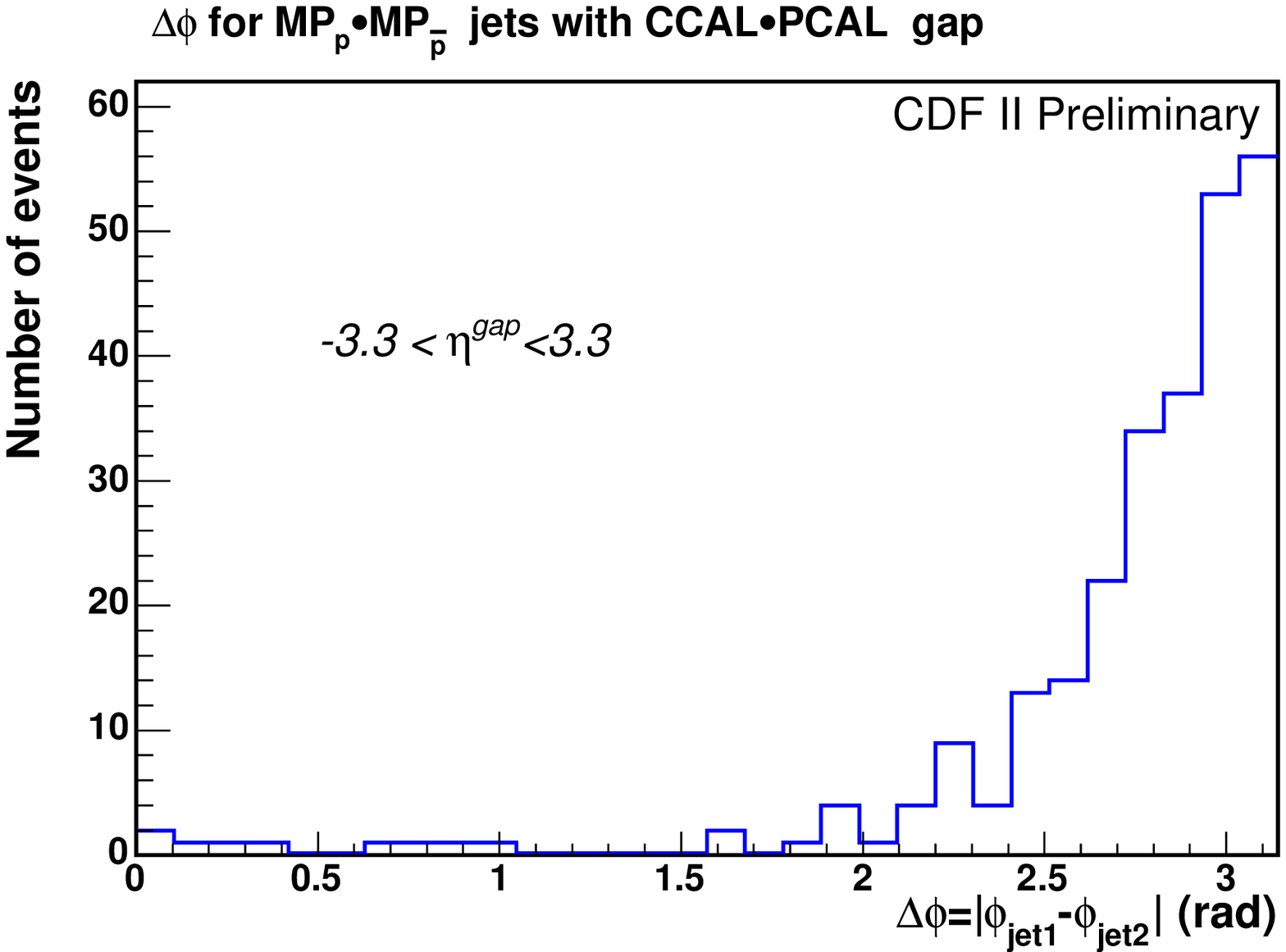}
\caption{
{\it Left:} Event rate gap fraction defined as $R_{gap}=N_{gap}/N_{all}$, for minimum bias (MinBias) and MP jet events with $E_T>2 (4)$~GeV;
{\it Right:} Azimuthal angle difference $\Delta\phi$ distribution of the two leading jets in a DD event with a central rapidity gap ($|\eta^{gap}|<3.3$).
}
\label{fig:forwardjets}
\end{center}
\end{figure}

An interesting process is dijet production in double diffractive (DD) dissociation.
DD events are characterized by the presence of a large central rapidity gap and are presumed to be due to 
the exchange of a color singlet state with vacuum quantum numbers.
A study of the dependence of the event rate on the width of the gap was performed using Run~I data with small statistics.
In Run~II larger samples are available.
Typical luminosities (${\cal L}\approx 1\div 10 \times 10^{31}$cm$^{-2}$sec$^{-1}$) 
during normal Run~II run conditions hamper the study of gap ``formation''
due to multiple interactions which effectively ``kill'' the gap signature.
Central rapidity gap production was studied in soft and hard diffractive events collected during a special 
low luminosity run (${\cal L}\approx 10^{29}cm^{-2}sec^{-1}$).
Figure~\ref{fig:forwardjets} (left) shows a comparison of the gap fraction rates, as function of the gap width (i.e. $\Delta\eta$) 
for minimum bias (MinBias), and MP jet events.
Event rate fraction is calculated as the ratio of the number of events in a given rapidity gap region divided by all events: $R_{gap}=N_{gap}/N_{all}$. 
The fraction is approximately 10\% in soft diffractive events, and approximately 1\% in jet events.
Shapes are similar for both soft and hard processes, and gap fraction rates decrease with increasing $\Delta\eta$.
The MP jets of gap events are produced back-to-back (Fig.~\ref{fig:forwardjets}, right).

\section{Exclusive production}

\begin{figure}[htp]
\begin{center}
\includegraphics[width=0.30\textwidth]{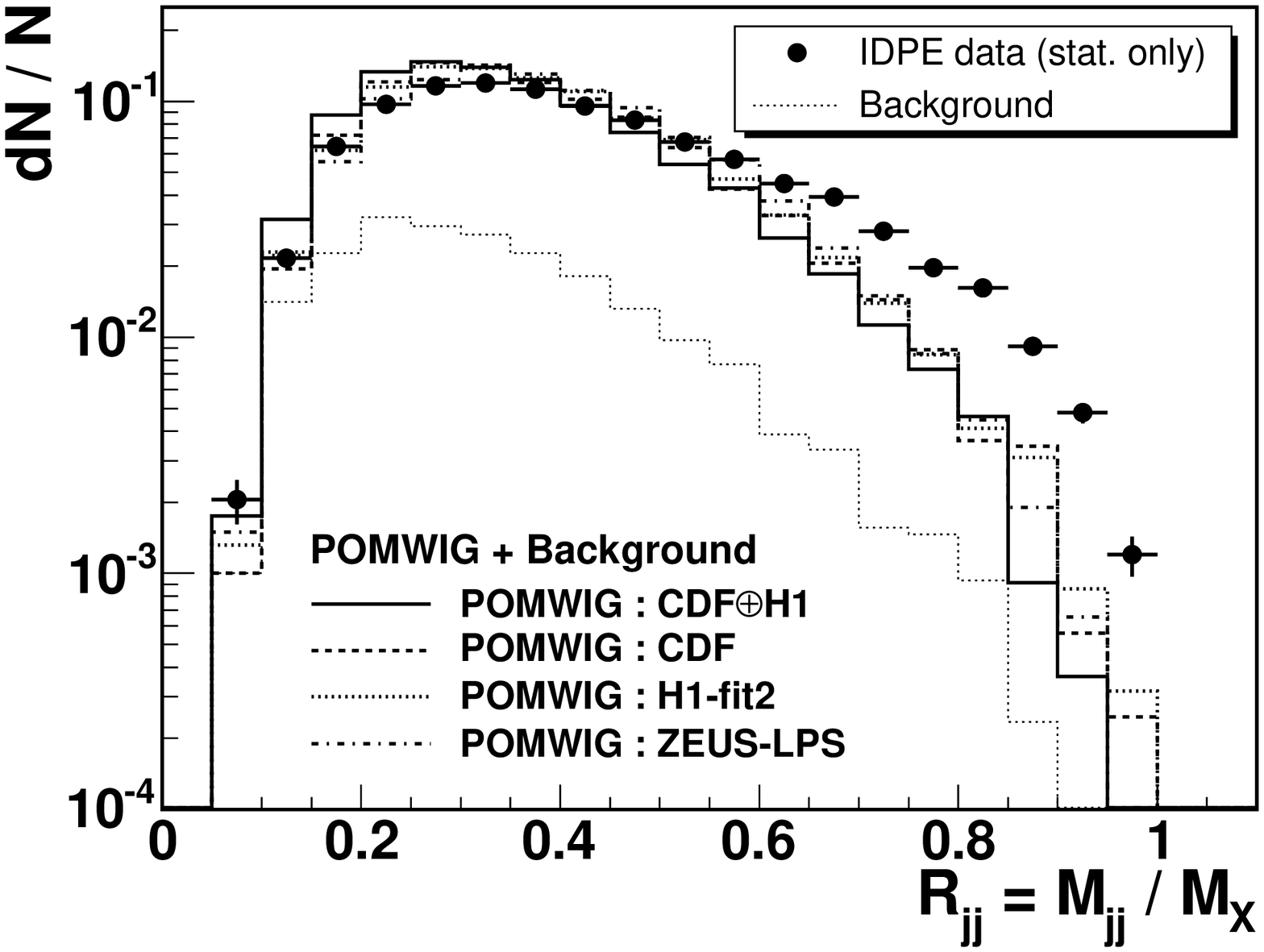}
\includegraphics[width=0.35\textwidth]{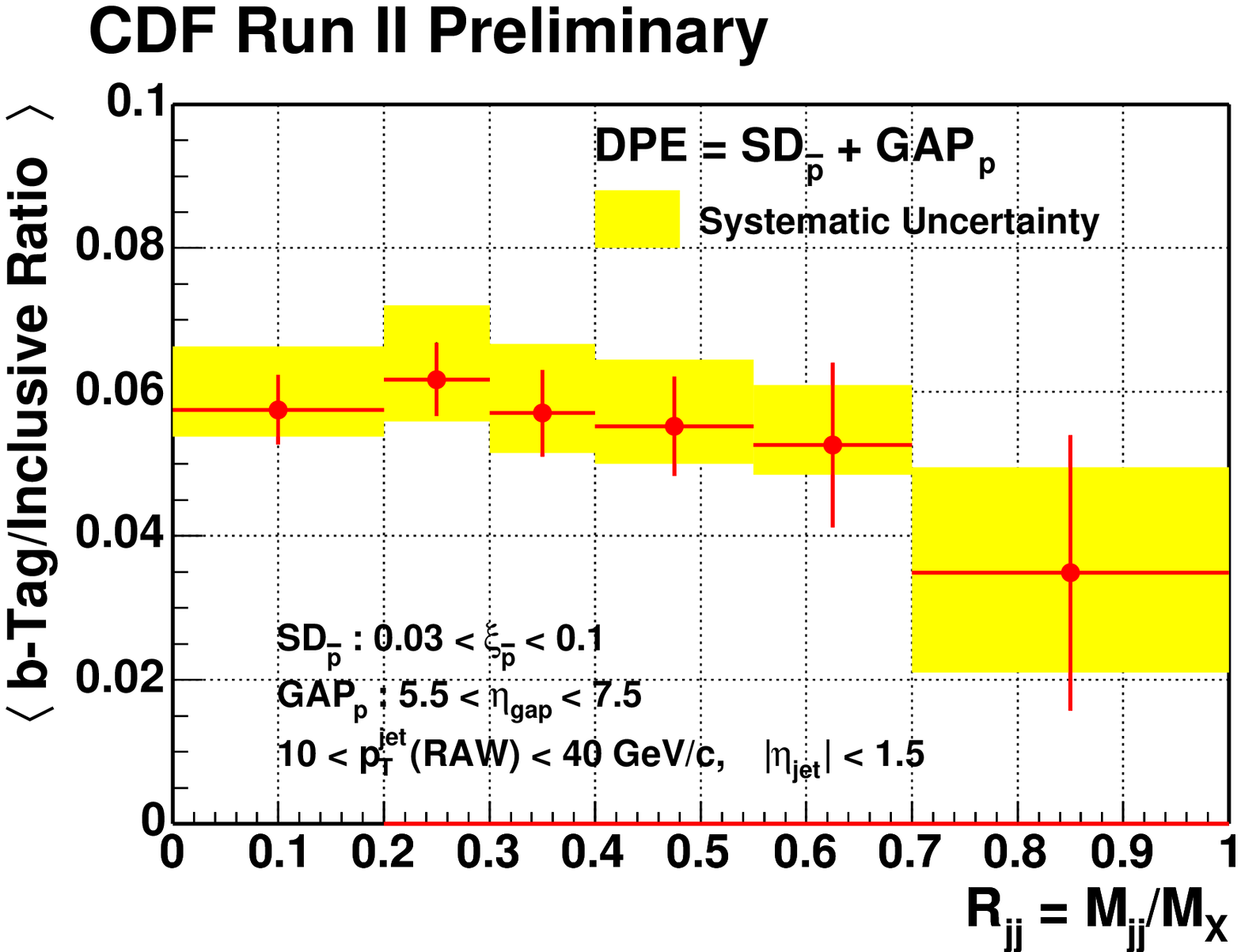}
\includegraphics[width=0.33\textwidth]{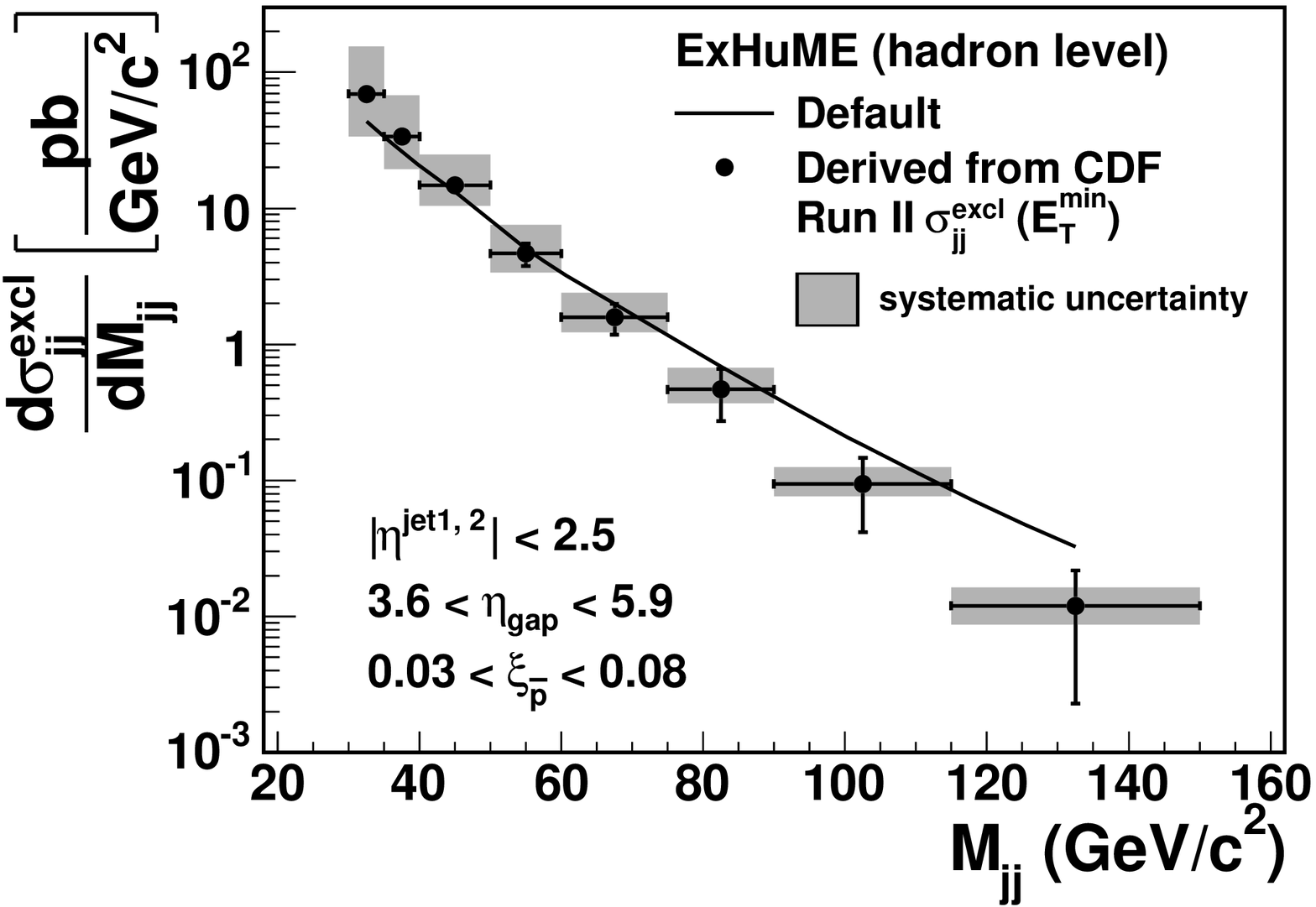}
\caption{{\it Left:} Dijet mass fraction $R_{jj}$ in inclusive DPE dijet data. 
An excess over predictions at large $R_{jj}$ is observed as a signal of exclusive dijet production;
{\it Center:} Ratio of b-tagged jets to all inclusive jets as a function of the mass fraction $R_{jj}$. 
The error band corresponds to the overall systematic uncertainty;
{\it Right:} The cross section for events with $R_{jj}>0.8$ is compared to predictions.}
\label{fig:exclusive}
\end{center}
\end{figure}

The first observation of the process of exclusive dijet production can be used as a benchmark to establish predictions 
on exclusive diffractive Higgs production, a process with a much smaller cross section\cite{kmr}.
A wide range of predictions was attempted to estimate the cross section for exclusive dijet and Higgs production. 
In Run~I, the CDF experiment set a limit on exclusive jet production~\cite{runI_excldijet}.
First observation of this process was made in Run~II.
The search strategy is based on measuring the dijet mass fraction ($R_{jj}$), 
defined as the ratio of the two leading jet invariant mass divided by the total mass calculated using all calorimeter towers. 
An exclusive signal is expected to appear at large $R_{jj}$ values (Fig.~\ref{fig:exclusive}, left).
The method used to extract the exclusive signal from the $R_{jj}$ distribution is based on fitting the data to MC simulations.
The quark/gluon composition of dijet final states can be exploited to provide additional hints on exclusive dijet production.
The $R_{jj}$ distribution can be constructed using inclusive or b-tagged dijet events. 
In the latter case, as the $gg\rightarrow q\bar{q}$ is strongly suppressed for $m_q/M^2\rightarrow 0$ ($J_z=0$ selection rule),
only gluon jets will be produced exclusively and heavy flavor jet production is suppressed. % (Fig.~\ref{fig:exclusive})
Figure~\ref{fig:exclusive} (center) illustrates the method that was used to determine the heavy-flavor composition of the final sample.
The falling distribution at large values of $R_{jj}$ ($R_{jj}>0.7$) indicates the suppression of the exclusive b-jet events.
The CDF result favors the model in Ref.~\cite{kmr2} (Fig.~\ref{fig:exclusive}, right).
Details can be found in Ref.~\cite{runII_excldijet}.

Exclusive $e^+e^-$ and di-photon production were studied using a trigger that requires forward gaps 
on both sides of the interaction point and at least two energy clusters in the electromagnetic calorimeters with transverse energy $E_T>5$~GeV. 
All other calorimeter towers are required to be below threshold.
In the di-electron event selection, the two tracks pointing at the energy clusters are allowed.
The CDF experiment reported the first observation of exclusive $e^+e^-$ production~\cite{excl_ee}.
A total of 16 $\gamma\gamma\rightarrow e^+e^-$ candidate events are observed, consistent with QED expectations.
Exclusive di-photon events can be produced through the process $gg\rightarrow \gamma\gamma$. Three candidate events were selected,
where one is expected from background sources (i.e. $\pi^0\pi^0$). 
A 95\%C.L. cross section limit of 410~pb can be set~\cite{excl_gg}, about ten times larger than expectations~\cite{kmr3}.

\section{Conclusions}

The results obtained  during the past two decades have led the way to the identification of
striking characteristics in diffraction. Moreover, they have significantly contributed to an understanding of
diffraction in terms of the underlying inclusive parton distribution functions.
The regularities found in the Tevatron data and the interpretations of
the measurements can be extrapolated to the LHC era.
At the LHC, the diffractive Higgs can be studied but not without challenges, as triggering and event acceptance
will be difficult.
Still, future research at the Tevatron and at the LHC holds much promise for further understanding of diffractive processes.

\section{Acknowledgments}
My warmest thanks to the organizers for the kind invitation and for a warm atmosphere, and in particular to S. Lami for providing the financial support.
%, and to those who contributed to the many results presented.


\begin{thebibliography}{0}
\bibitem{rapidity}
The terms rapidity and pseudorapidity are used interchangeably.
The pseudorapidity $\eta$ is defined as $-\ln \tan(\theta/2)$, where $\theta$ is
 the polar angle with respect to the proton beam direction.
The azimuthal angle is denoted by $\phi$, and the transverse energy of a jet, $E_T$, is defined as $E_T\equiv E\cdot \sin\theta$.
\bibitem{xbj}
For each event, $x_{Bj}$ is evaluated from the $E_T$ and $\eta$ of the jets using the equation
$x_{Bj}=\frac{1}{\sqrt{s}}\sum_{i=1}^3E_T^ie^{-\eta^i}$.
\end{thebibliography}
\end{document}